\documentstyle[twoside,fleqn,espcrc2,epsf]{article}

\begin{document}

\title{On the Parity-Flavor-Breaking Phase
in QCD With Two Flavors of Wilson Fermions}

\author{Khalil M.~Bitar\address{Supercomputer Computations Research
    Institute,
    Florida State University,\\
    Tallahassee,  Florida 32306-4130, USA\\
    {\tt kmb@scri.fsu.edu}}\thanks{This research was supported by the
    U.S. Department of Energy through Contract Nos. DE-FG02-92ER40979
    and DE-FG05-85ER250000.}}

\begin{abstract}
  We present further data in search of a parity-flavor-breaking phase
  in simulations of dynamical QCD with two flavors of light Wilson
  fermions in the strong coupling region. This is done on lattice
  sizes of $4^4$ an up to $10^4$ for a variety of values of $\beta$
  and $\kappa$ as well as the coefficient, $h$, of an explicit
  breaking term included in the action.  We confirm the existence of a
  region in the $\beta-\kappa$ plane where such a phase exists at
  $\beta=6/g^2$ smaller than 5.0 and above $\kappa_c$.
\end{abstract}

\maketitle

\section{Introduction}
\label{introduction}
We extend our search for a parity-flavor-breaking phase in simulations
of dynamical QCD with two flavors of light Wilson fermions to the
strong coupling region of $\beta$ less than 5.0.

The signature used in this search is the same used in our previous
work ~\cite {bitar97} . We introduce an explicit breaking term into
the action with a small coefficient and then study the finite size
behavior of its expectation value as this coefficient tends to zero.

\section{Numerical Simulations}
\label{Numerical Simulations}

We report here on simulations done with two flavors of Wilson fermions
at $\beta=3.0, 3.5, 3.7 $ and $4.0$ on volumes of $4^4$, and up to
$10^4$ for a variety of values of $\kappa$ ranging from less than the
appropriate $\kappa_c$ to values greater than $\kappa_c$. Some of
these simulations (denoted by the letter $C$ in the tables) are
continuing and simulations at $\beta=4.5$ are in progress.

We introduce into the QCD action a term of the form $i h \bar{\psi}
\gamma_5 \tau_3 \psi$ where $\tau_3$ is a $2\times2$ matrix
representing the third element of the generators of flavor SU(2)
algebra.  Upon integrating the fermionic variables this is reflected
in the simulation by the product of two determinants: $Det M(h) * Det
M(-h) $ where $M(h)$ is given by a simple modification of the Wilson
Matrix $M_w$ as:
   $$ M(h) = M_w + ih\gamma_5$$
As pointed out by Aoki  \cite{ref2}, we also have here 
   $$ \gamma_5 M(-h) \gamma_5 = M^\dagger $$
and:
   $$ Det M(-h) = Det M^\dagger (h)\, . $$
   
   Simulations were done for the parameter $h$ taking values ranging
   from 0.005 to 0.1. For the volume dependence we concentrate on the
   smaller values of $h$ and in particular $ h= 0.005$ for all three
   volumes considered and mostly for values of $\kappa$ greater than
   $\kappa_c$.

   The order parameter we compute is the expectation value of the
   operator $ i \bar{\psi} \gamma_5 \tau_3 \psi$. With our notation
   this is given as

$$ PF= - Im Tr(\gamma_5 M^{-1}(h)) $$  

\section{Results}
\label{Results}

For the values of $\beta$ considered, simulations were performed, as
mentioned above, at various values of $\kappa$ both below and above
$\kappa_c$. We shall present the data and results for each value of
$\beta$ considered separately.  The results are shown in Tables~1--4.

\setlength{\tabcolsep}{4.5pt}
\begin{table}[t]
\caption{Parameters and measured order parameter $PF_L$ for the
 case of $\beta=3.0$ on lattices of volume $L^4$,  for $L=4$,$6$,  and
$10$.\label{tabbeta3.0} }
\begin{center}
\begin{small}
    \begin{tabular}{ccccc}
      \hline
       $\kappa$
 & $h$ & $PF_4$ & $ PF_6$ & $PF_{10}$\\
      \hline
       0.1900 & 0.005 &             &0.159(3  )   & 0.159(1)  \\
       0.1900 & 0.050 &             &1.503(24)    & 1.503(8)    \\
       0.1900 & 0.100 &             &2.670(37)    &             \\
      \hline
       0.220  & 0.005 &             &0.54 (4)     & 0.526 (14)  \\
       0.220  & 0.050 &             &2.681(73)    & 2.690 (26)  \\
       0.220  & 0.100 &             &3.654(64)    &             \\
      \hline
       0.250 & 0.005  & 2.13 (64)   & 2.60 (30)   & 2.99 (13) C \\
       0.250 & 0.050  &             & 3.67 (11)   &             \\
       0.250 & 0.100  &             & 4.232(76)   &             \\
      \hline
    \end{tabular}
\end{small}
\end{center}
\end{table}
\begin{table}
\caption{Parameters and measured order parameter $PF_L$ for the
 case of $\beta=3.5$ on lattices of volume $ L^4$,  for $L=4, 6, 10 $.
\label{tabbeta3.5}}
\begin{center}
\begin{small}
    \begin{tabular}{cccccc}
      \hline
       $\kappa$ & $h$ & $PF_4$ & $PF_6$ & $PF_{10}$\\
      \hline
       0.1900 & 0.005 &   &0.1648(33)   & 0.1639(11) \\
       0.1900 & 0.050 &   &1.531 (26)   & 1.533(9)   \\
       0.1900 & 0.100 &   &2.682 (38)   &             \\
      \hline
       0.240 & 0.005  & 1.42 (60) &2.16(28)C  & 2.43(9)C   \\
       0.240 & 0.050  &           &3.171(94)  &       \\
       0.240 & 0.100  &           &3.848(79)  &       \\
      \hline
       0.250 & 0.005  & 0.68(42)  &0.354(5)C  &     \\
      \hline
    \end{tabular}
\end{small}
\end{center}
\end{table}
\begin{table}[t]
\vspace{-14pt}
  \caption{Parameters and measured order parameter $PF_L$ for the
 case of $\beta=3.7$ on Lattices of Volume $ L^4$,  with $L=4, 6, 8,10$.
\label{tabbeta3.7}}
\begin{center}
\begin{small}
    \begin{tabular}{lllll}
      \hline
      \multicolumn{1}{c}{$\kappa$} &
        \multicolumn{1}{c}{ $h$}   &
         \multicolumn{1}{c}{$PF_4$} &
          \multicolumn{1}{c}{$PF_6 $}  &
          \multicolumn{1}{c}{  $PF_{10}$}\\[.5ex]
      \hline
       0.2300 & 0.005 &1.14 C(45) & 1.78(26)C&  1.8(2)C \\
       0.2400 & 0.005 &0.673(380) & 1.128(239)&  2.43(10)C  \\[1ex]
      \hline
\noalign{\vspace{4pt}}
          & & & &   
          \multicolumn{1}{c}{  $PF_{8}$}\\[.5ex]
       \hline
       0.2500 & 0.005 &0.337(193) & 0.360(62) & 0.355(63)   \\
      \hline
    \end{tabular}
\end{small}
\end{center}
\end{table}

\begin{table}
  \caption{Parameters and measured order parameter $PF_L$ for the
 case of $\beta=4.0$ on Lattices of Volume $ L^4$,  with $L=4, 6, 10$.
\label{tabbeta4.0}}
\begin{center}
\begin{small}
    \begin{tabular}{ccccc}
      \hline
       $\kappa$ & $h$ & $PF_4$ & $ PF_6$ & $PF_{10}$\\
      \hline
       0.1900 & 0.005 &  &0.1715 (38)  & 0.1719(13) \\
       0.1900 & 0.050 &  &1.574  (31)  & 1.569 (10) \\
       0.1900 & 0.100 &  &2.691  (38)  &            \\
      \hline
       0.2200 & 0.005 & 0.58(21) &1.20  (18) & 1.25(07)C  \\
       0.2200 & 0.050 &          &2.605 (76) &            \\
       0.2200 & 0.100 &          &3.358 (64) &            \\
      \hline
       0.2300 & 0.005 &0.18(03)  &0.33C(11) &             \\
      \hline
       0.2400 & 0.005 &0.28(13)  & 1.30(23)C  &  2.38(10)C \\
       0.2400 & 0.050 &          & 2.459 (86) &            \\
       0.2400 & 0.100 &          & 3.517 (68) &            \\
      \hline                     
\end{tabular}
\end{small}
\end{center}
\end{table}

\begin{figure}[hbt]
\vspace{4in}
\includegraphics{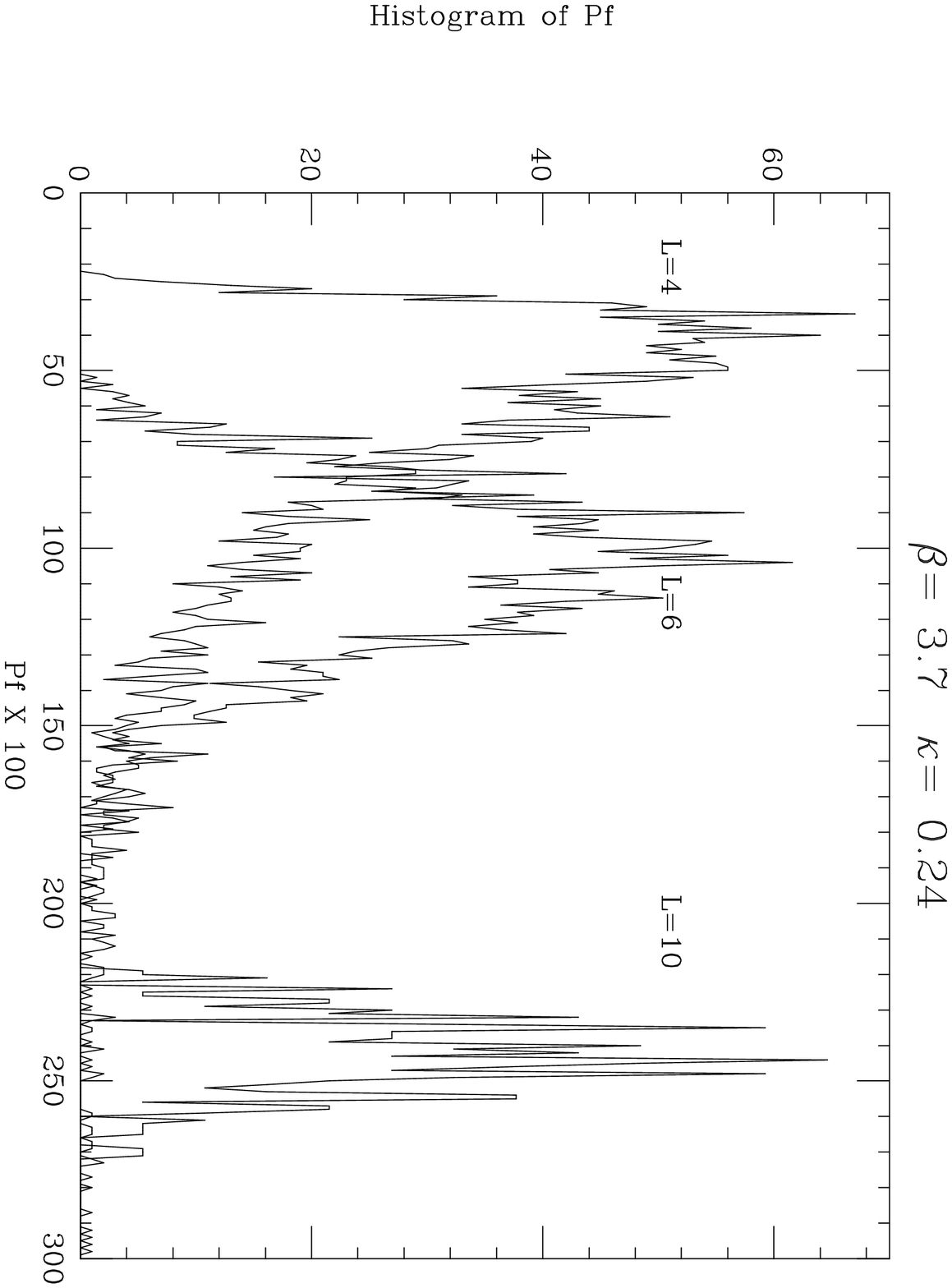}
\includegraphics{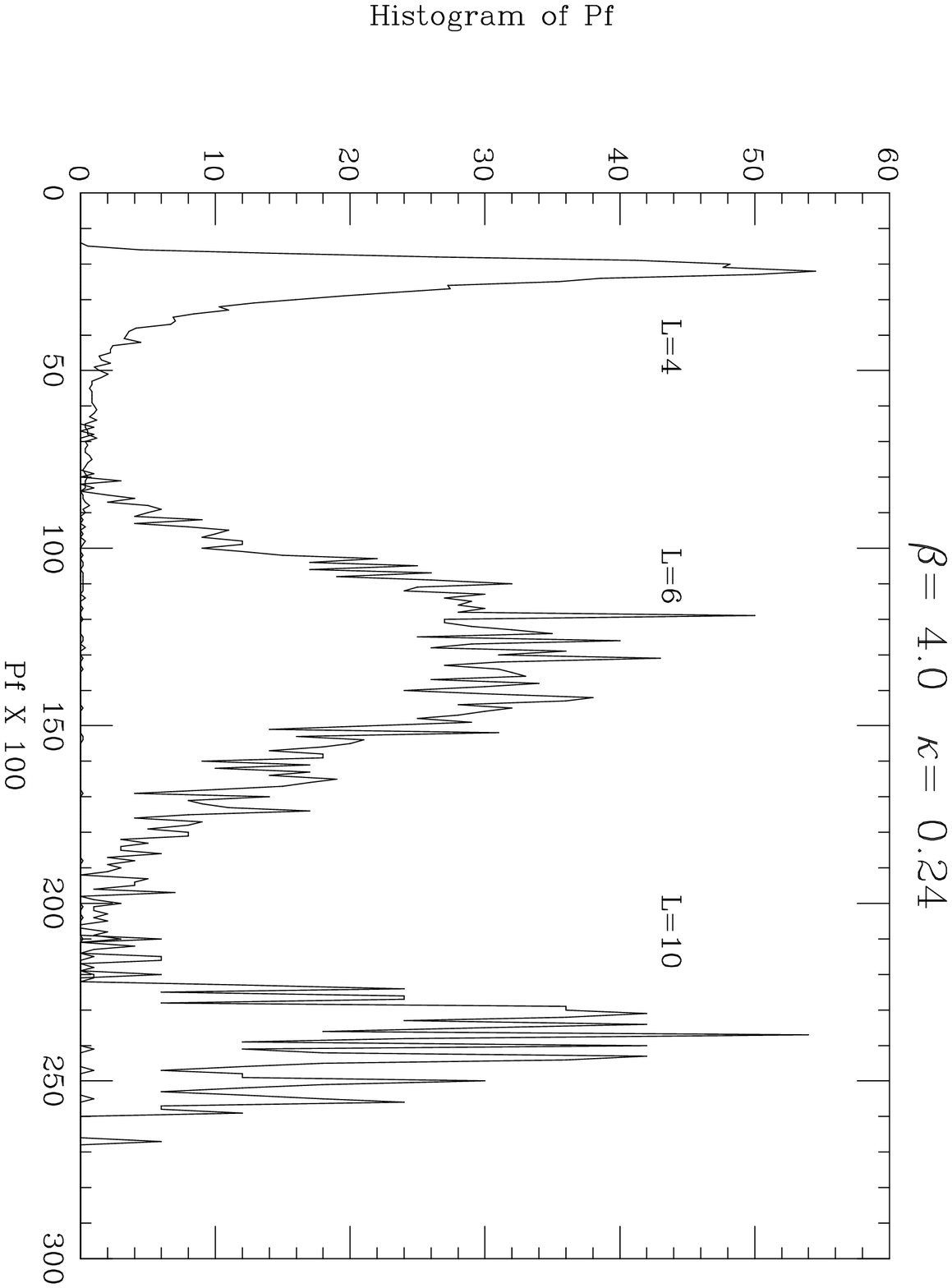}
\caption{Histogram signature of a parity-flavor-breaking phase for
$\beta=3.7$ and $4.0$; $\kappa=0.24$.\label{pfhisto}}
\end{figure}

It is clear that if one concentrates on the behavior with volume of
the simulations at $h=0.005$, one may quickly note that a signature for
the parity-flavor-breaking phase is indicated for the following values
of $\beta $ and $\kappa$: (3.0, 0.25), (3.5, 0.24), (3.7, 0.24) and
(4.0, 0.24).  See Fig.~\ref{pfhisto} for the latter two points.
 
This is also confirmed by following the behavior of the order
parameter as a function of $\kappa$ at constant volume and constant
$\beta$. it is clear that all values above the order parameter goes
through a relative maximum.

The behavior of the order parameter at all other values of $\beta$ and
$\kappa$, computed in Tables~1--4, are consistent with an absence for
a signature of a parity-flavor-breaking phase. This behavior is in
fact very similar to the behavior for values of $\beta \geq 5.0$ which
were discussed in ref.~\cite{bitar97}.

It is important to note here that the phase is not indicated for some
points at the same $\beta$ but larger values of $\kappa$.  In
particular the phase disappears at (3.5, 0.25), at (3.7, 0.25) .
 
\begin{figure}[t]
\vspace{2.0in}
\includegraphics{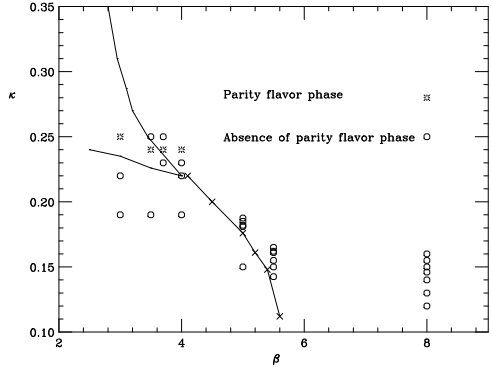}
\caption{Parity-flavor phase diagram. Continuous line represents 
  approximate positions for chiral and finite temperature 
  transitions at $N_t=4$ for comparison.\label{parphase}}   
\end{figure}
It is clear from the discussion above that QCD with two flavors of
Wilson fermions does exhibit a parity-flavor-breaking phase at values
of $\beta$ up to 4.0 , as postulated by Aoki and collaborators. It has
also been demonstrated in ~\cite{bitar97} that this phase does not
seem to extend beyond $\beta=5.0$ . This information is exhibited in
Fig.~\ref{parphase}.  It is then possible that this phase pinches out,
in a manner similar to that in the NJL model, at $\beta < 5.0$. In
this case, this phase would not be relevant for the discussion of the
approach to the chiral limit in QCD and the ensuing Goldstone nature
of the pions for $\beta > 5.0$. In fact, all indications are such
that, as shown formally sometime ago, this is simply related to the
approach to zero lattice spacing and infinite volume. 
\vfil\eject

\section*{Acknowledgements}

I wish to thank Urs Heller and Robert Edwards for providing a modified
QCD code for performing the simulations reported in this paper. All
these simulations were done on the SCRI IBM compute cluster.

\bibliographystyle{unsrt}

\end{document}